\documentclass{aa}

\usepackage{graphicx}
\usepackage[varg]{txfonts}
\usepackage{natbib}
\usepackage{lscape} 

\bibpunct{(}{)}{;}{a}{}{,} 

\def \FLL {\emph{Fermi}-LAT~}

\begin{document}

\title{Twelve-year update of the MST catalogue of $\gamma$-ray source candidates above 10 GeV 
       and at Galactic latitudes higher than 20\degr}

\author{R. Campana
                \inst{1},
                E. Massaro
                \inst{2}
}
\institute{INAF/OAS, via Piero Gobetti 101, I-40129 Bologna, Italy
\and INAF/IAPS, via Fosso del Cavaliere 100, I-00133 Roma, Italy
}           
\offprints{{riccardo.campana@inaf.it} The catalogue is only available at the CDS via anonymous ftp to {cdsarc.u-strasbg.fr} (130.79.128.5) or via {http:// cdsarc.u-strasbg.fr/viz-bin/qcat?J/A+A/vol/page}}
\date{Received ..., accepted ...}
\markboth{R. Campana, E. Massaro}
{12 year update of the MST catalogue of $\gamma$-ray source candidates}

%
\abstract{
We present an updated version catalogue of $\gamma$-ray source candidates, 
12Y-MST, selected using the minimum spanning tree (MST) algorithm on the 
12-year \FLL sky (Pass 8) at energies higher than 10~GeV. 
The high-energy sky at absolute Galactic latitudes above 20\degr\ has been 
investigated using rather restrictive selection criteria, resulting in a 
total sample of 1664 photon clusters, or candidate sources. 
Of these, 230 are new detections, that is, candidate sources without any association in 
other $\gamma$-ray catalogues. 
A large fraction of them have interesting counterparts, most likely blazars.
We describe the main results on the catalogue selection and search of 
counterparts.

We also present an additional sample of 224 candidate sources (12Y-MSTw), which are clusters
that we extracted by applying weaker selection criteria: about 57\% of them have not been reported 
in other catalogues.
}
%
\keywords{      Gamma rays: general --
                        Gamma rays: galaxies --
                        Methods: data analysis 
                 }

\authorrunning{R. Campana, E, Massaro}
\titlerunning{12Y-MST, the 12-year update of the 9Y-MST catalogue of $\gamma$-ray source candidates}

\maketitle


\section{Introduction} \label{s:intr}

In a previous paper \citep{campana18} we reported a complete catalogue 
of $\gamma$-ray source candidates at Galactic latitudes $|b| > 20\degr$, selected by means
of the minimum spanning tree (MST) clustering algorithm, 
which was applied to extract
significant photon spatial over-densities in \emph{Fermi}-Large Area Telescope (LAT) mapping of the sky at 
energies higher than 10 GeV \citep{campana08,campana13}.
In that work we examined the data accumulated on the full sky, over the nine-year time interval 
from 2008 August 4 to 2017 August 4, excluding a 40\degr\ wide belt around the Galactic 
Equator where the high-energy emission is strongly affected by many sources and extended 
structures.
This 9Y-MST catalogue (hereafter 9Y) contains 1342 $\gamma$-ray candidate sources.

In several other papers, the MST cluster-finding algorithm was applied to identify several 
new blazar candidates \citep{bernieri13,campana15,campana16a,campana16b,campana16c,campana17} 
and also to study the field of the Large Magellanic Cloud (LMC), allowing the first 
detection of high-energy emission from the supernova remnants N 49B and N 63A \citep{campana18b}.
In general, it was shown that the advantage of MST 
is its capability of extracting `noise' clusters from the background with even a low number of photons, 
and thus to discover rather faint high-energy sources.

In this paper we report an update of the 9Y-MST catalogue that is based on a longer observational 
run covering three additional years of data. 
Using essentially the same approach, we compiled a new catalogue of $\gamma$-ray sources
and candidates (12Y-MST) that includes 322 more entries than 9Y for a total of 1664 clusters.
Throughout this paper, the term `photon cluster', or simply `cluster', is used interchangeably to stand for $\gamma$-ray source or candidate source.

In Section \ref{s:mst_descr} we summarise the main parameters we used for cluster 
selection. In Section \ref{s:FLdata} we describe the data reduction. The MST analysis and the 
general properties of the 12Y-MST catalogue are described in Section \ref{s:cat12y}. 
We also present an additional catalogue of 224 candidate sources, obtained by applying 
less severe selection criteria (12Y-MSTw). This is useful to compare our method with 
other source-detection tools.
The content and the characteristics of this sample are described in Section
 \ref{s:corr12y}, and in Sections \ref{s:newblz} and \ref{s:sumcon} 
we summarise and discuss our results.

\section{MST cluster detection and main parameters}
\label{s:mst_descr}

The MST (see e.g. \citealt{cormen09} and also
\citealt{campana08,campana13}) is a topometric tool that can be used to search for 
spatial concentrations in a field of points. 
As previously stated, we have applied this method to the $\gamma$-ray sky, 
and detailed descriptions of the MST and of the selection criteria were presented 
elsewhere (e.g. in \citealt{campana18}). We therefore describe here only the meaning 
and the relevance of the main parameters we used for the cluster selection that is reported 
in the catalogue.

The MST creates the unique tree  
that connects all the points in a two-dimensional 
metric space on the condition that the sum of the (weighted) distances $\{\lambda_i\}$ 
of all pairs of connected points is minimum, provided that all distances are
different.
We considered the angular distances of the incoming direction of the photons.
Clusters are obtained by eliminating all segments in the tree with a length 
$\lambda > \Lambda_\mathrm{cut}$.
This so-called separation value, suitably chosen, is defined in units of the mean edge 
length in the region in which the tree is computed $\Lambda_m = (\Sigma_i \lambda_i)/N$, 
with $N$ equal to the total number of points.
A set of disconnected sub-trees is thus obtained; then, an elimination process removes all the sub-trees with a number of nodes $n \leq N_\mathrm{cut}$, leaving only the 
clusters that have a size higher than a fixed threshold. 
The remaining set of sub-trees provides a first list of candidate clusters, and a 
secondary selection is applied to extract the most robust candidates as $\gamma$-ray 
sources.
According to \cite{campana13}, this procedure is based on some useful parameters, such as
the clustering parameter,
\begin{equation}
g_k = \Lambda_m / \lambda_{m,k}
,\end{equation}
with $\lambda_{m,k}$ the mean distance of points in the $k$-th cluster. 
Another quantity that is very useful for assessing the significance of the surviving clusters 
is the 
cluster magnitude,
\begin{equation}
M_k = n_k~ g_k  
,\end{equation}
where $n_k$ is the number of nodes in the cluster $k$. 
The probability of obtaining a given magnitude value combines that of selecting a cluster with
$n_k$ nodes together with its `clumpiness', compared to the mean separation in the field.  
It was found that $\sqrt{M}$ is a good estimator of statistical significance of MST clusters.
In particular, a lower threshold value of $M$ in the range 15--25 would reject the large 
majority of spurious (low-significance) clusters. 
For each cluster, the {centroid coordinates} are obtained by means of a 
weighted average of the photon coordinates.
Additionally, the {median radius} $R_{m,k}$ is defined as the radius of 
the circle centred at the centroid and containing $n_k/2$ photons, and the {maximum radius} 
$R_{max,k}$ is equal to the angular distance between the centroid and the farthest photon in the cluster.
For clusters associated with genuine point-like sources, $R_{m,k}$ should be smaller than or 
comparable to the 68\% containment radius of instrumental point spread function 
\citep[PSF, see][]{ackermann13b}, while the maximum radius gives information about the total 
extension of the cluster.
For more details, see \cite{campana13} and the previous 9Y paper \citep{campana18}.

\section{\emph{Fermi}-LAT dataset} \label{s:FLdata}

The full \emph{Fermi}-LAT dataset was downloaded in the form of weekly files from the FSSC 
archive\footnote{\url{http://fermi.gsfc.nasa.gov/ssc/data/access/}}. It includes events above 10 GeV in the 12 years from 2008 
August 4 to 2020 August 4 that were processed with the Pass 8, release 3 reconstruction algorithm 
and responses. 
The event lists were then filtered by applying the standard selection criteria on data quality 
and zenith angle (source class events, \texttt{evclass} = 128, front and back converting, \texttt{evtype} = 3, 
up to a maximum zenith angle of $90\degr$). 
Events were then screened for standard good time interval selection.

For the sake of consistency with the previous analysis, the MST algorithm was applied to the data 
in the sky regions with $|b| > 18\degr$ because clusters in the Galactic belt with a small number 
of photons are not stable, that is, their number of photons and other parameters change even for 
small variations in $\Lambda_\mathrm{cut}$.
The final dataset contains 324\,133 photons, about 53\% and  
 47\% of which are in the north and south Galactic regions, respectively. This difference is 
mainly due to the nonuniform spatial distribution of the exposure, as already noted for 
the 9Y-MST catalogue.
The detection of photon clusters by means of MST depends upon the mean spatial density of events, 
which increases largely moving from the Galactic poles to the equator.

\begin{figure}[ht]
\vspace{0.1 cm}
\centerline{
\includegraphics[width=\columnwidth]{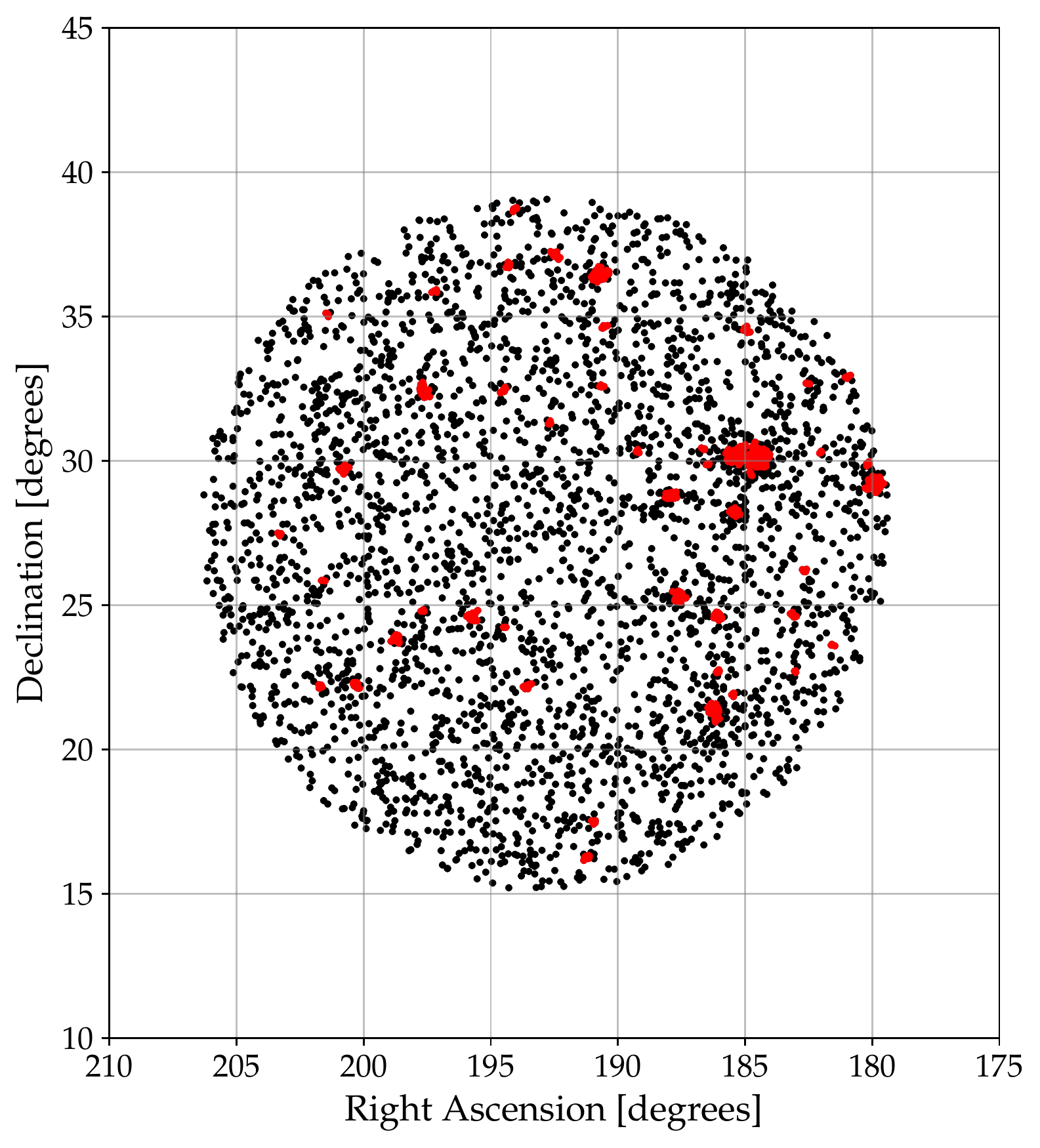}}
\caption{Map of photons at energies higher than 10 GeV in a region centred at the north
Galactic pole and with a radius of 12$^\circ$. 
Individual photons are represented by black points, and those in the clusters after the 
primary selection are shown in red.  
The two close large clusters correspond to the well-known BL Lac objects 5BZB J1217+3007
(ON 325) and 5BZB J1221+3010 (1ES 1218+304). }
\label{f:a}
\end{figure}

\subsection{Primary selection}\label{ss:prisel}

The division of the sky into several regions followed the same procedure as in 9Y.
All the regions were quite broad to reduce the effect of very rich clusters
on $\Lambda_m$, but the increase in photon density due to the longer exposure
required the use of the lower separation distance $\Lambda_\mathrm{cut} = 0.6\,\Lambda_m$ 
instead of $0.7\,\Lambda_m$ that was used in the 9Y, while $N_\mathrm{cut}$ remained 
equal to 3.
The resulting lists of clusters in the various regions were merged together and a 
first preliminary set of 4841 clusters was obtained, including double 
detections in the overlapping strips.
The results of this primary selection and the properties of the search regions
are summarised in Table~\ref{t:regions}.

To illustrate this primary selection, Figure~\ref{f:a} shows an image of the 
photon field around the north Galactic pole with the detected clusters. There are two close 
large clusters, but many others have far fewer photons.

The criteria adopted in the primary selection are generally weak, and it cannot be
excluded that a fraction of clusters with a low number of photons originated 
from statistical fluctuations of the background photon density or were caused by the occurrence 
of extended local enhancements, and therefore they cannot be related to genuine 
$\gamma$-ray sources.
The application of stronger selection criteria is then necessary to reduce
the number of these spurious detections.
Nevertheless, making this further selection can exclude some true clusters (i.e., 
clusters that are really associated with sources), and therefore a reasonable trade-off must be taken 
into account.

Several types of spurious clusters are possible.
Photon density fluctuations are more relevant for clusters with a low number of 
photons, for example, four or five, which is just above the elimination threshold.
These clusters can be found in the surroundings of particularly rich clusters
(typically with $n$ above a few hundred) associated with bright sources.
As we discuss in Section~\ref{ss:sat}, these rich clusters have  
$R_\mathrm{max}$ in the range between $\sim$20\arcmin\ and $\sim$40\arcmin, and the 
possibility of finding one or two clusters that are disconnected from the rich cluster by 
a single edge with a length just above $\Lambda_\mathrm{cut}$ is not negligible.
These structure are called {satellite clusters}.

Other spurious features can be found in the regions in which the local photon density is slightly
higher than the mean density, as is observed when the Galactic belt is approached. 
In these regions some clusters with a low density (below the selection
threshold) might be connected by one or two background photons and thus build a 
structure with a number of photons $\sim$10 and $g$ typically lower than 3 (in a 
few cases lower than 2), but they produce a value of $M$ above the threshold and have a 
spatial extension higher than 10\arcmin.
These {bridged clusters} (Section~\ref{ss:bri}), however, cannot simply be ruled out 
because they can include a small sub-cluster that corresponds to a genuine source, and further 
analyses are useful to determine their nature.

\subsection{Secondary selections}
\label{ss:secsel}

As we did for the 9Y catalogue, stronger `superselection' criteria were applied
for sorting clusters with high significance.  
This procedure has been proved to be efficient in accounting for the background non-uniformities, 
increasing the probability of selecting clusters that correspond to genuine candidate $\gamma$-ray 
sources.
These criteria consist in different threshold values on $M$ and $g$, proportional to
the mean photon density.
The threshold values were chosen the same for the 9Y, that is, 
for high latitudes and poles, $|b| > 50^\circ$: $M>18$ or $g >3.5$; for middle latitudes, $30^\circ <|b|< 50^\circ$: $M>20$ or $g>4.0$;
for the  external peri-Galactic belt, $20^\circ <|b|< 30^\circ$ with $0^\circ< l <330^\circ$: $M>22$ or $g>4.2$;
for the central  peri-Galactic belt, $20^\circ <|b|< 30^\circ$ with $-30^\circ< l <30^\circ$: $M>24$ or $g>5.0$;

The alternative (or) above is intended in the Boolean sense, that is, 
we selected clusters with either $M>18$ or $g >3.5$ for the Galactic poles.
For some clusters, a more detailed analysis is necessary 
either to avoid spurious features or to resolve coupled structures, as
discussed in Section~\ref{s:sat12y}.
A new final set of 1664 clusters was thus obtained, which are all included in the 12Y-MST 
catalogue. These are 322 clusters more than in 9Y.

The thresholds quoted above are rather severe and reject the majority of spurious 
clusters, but they might also filter out some real features related to genuine sources, 
particularly in sky regions with low background.
To obtain a more complete listing of clusters, we performed another secondary selection 
with slightly lower threshold values, except in the fourth region closer to the 
Galactic centre. 
The parameters are:
for the igh latitudes and poles, $|b| > 50^\circ$: $M>15$ or $g >3.0$;
for the middle latitudes, $30^\circ <|b|< 50^\circ$: $M>18$ or $g>3.5$;
for the external peri-Galactic belt, $20^\circ <|b|< 30^\circ$ with $0^\circ< l <330^\circ$: $M>20$ or $g>4.0$;
for the central  peri-Galactic belt, $20^\circ <|b|< 30^\circ$ with $-30^\circ< l <30^\circ$: $M>24$ or $g>5.0$.

The application of these {weak} (w) selection criteria resulted in an additional 
sample of  224 clusters (called 12Y-MSTw). This increased the total number to 1888.

The numbers of candidate sources in each region resulting from both the {standard} (s) and the 
{weak} (w) selection are also given in Table~\ref{t:regions}.

\begin{table*}[ht]
\centering
\caption{Sky regions used in the MST search for photon clusters.
There is an overlap at each side of the regions. 
The total number of photons, the solid angle of the region for a single hemisphere, and 
the mean angular separation lengths for the north and south Galactic regions are given; 
the clearly evident N-S asymmetry is mainly due to the different exposures. 
In the last three columns we report the number of clusters found in each region with
the primary selection and after the application of the secondary standard and weak selections, 
as described in the text.
The Galactic longitude intervals are A:~$0\degr < l < 182\degr$, B:~$178\degr < l < 362\degr$, 
C:~$0\degr < l < 41\degr$, D:~$39\degr < l < 182\degr$, E:~$178\degr < l < 331\degr$, and 
F:~$329\degr < l < 362\degr$.}\label{t:regions}
\begin{tabular}{ccrrcccccccccc}
\hline
Region                          &  Solid angle    &     \multicolumn{2}{c}{Photon number}   & \multicolumn{2}{c}{$\Lambda_\mathrm{m}$} & \multicolumn{2}{c}{$\Lambda_\mathrm{cut}$} & \multicolumn{2}{c}{$N_\mathrm{ps}$}  & \multicolumn{2}{c}{$N_\mathrm{s}$} &  \multicolumn{2}{c}{$N_\mathrm{w}$} \\
                             &     sr          &   N   &    S   &     N     &     S     &     N     &     S     &  N  &  S  &  N  &  S  &  N  &  S  \\
\hline
$|b| > 78\degr$              &    0.1373       &  5083 &  3154  &  9\farcm5 & 18\farcm7 &  5\farcm7 & 11\farcm2 &  46 &  48 &  33 &  41 &  37 &  44 \\
\hline
$64\degr < |b| < 81\degr$ A  &    0.2824       & 11729 & 6299  &  8\farcm7 & 14\farcm2 &  5\farcm2 &  8\farcm5 &  95 &  99 &  69 &  65 &  79 &  74 \\  
$64\degr < |b| < 81\degr$ B  &    0.2855       & 10034 & 5875  & 10\farcm0 & 15\farcm0 &  6\farcm0 &  9\farcm0 &  93 &  99 &  68 &  57 &  77 &  71 \\ 
\hline
$49\degr < |b| < 66\degr$ A  &    0.5045       & 18939 & 14109  &  9\farcm8 & 12\farcm1 &  5\farcm9 &  7\farcm3 & 176 & 202 & 111 & 110 & 128 & 129 \\  
$49\degr < |b| < 66\degr$ B  &    0.5101       & 16820 & 13423  & 10\farcm7 & 12\farcm6 &  6\farcm4 &  7\farcm5 & 176 & 191 & 104 & 122 & 123 & 138 \\
\hline
$34\degr < |b| < 51\degr$ A  &    0.6923       & 27127 & 24953  & 10\farcm1 & 11\farcm0 &  6\farcm1 &  6\farcm6 & 332 & 333 & 143 & 126 & 171 & 154 \\  
$34\degr < |b| < 51\degr$ B  &    0.6999       & 23800 & 23396  & 11\farcm3 & 11\farcm2 &  6\farcm8 &  6\farcm7 & 402 & 333 & 150 & 141 & 169 & 164 \\
\hline
$18\degr < |b| < 36\degr$ C  &    0.1995       & 14277 & 12031  &  8\farcm1 &  8\farcm7 &  4\farcm8 &  5\farcm2 & 219 & 191 &  23 &  26 &  26 &  28 \\
$18\degr < |b| < 36\degr$ D  &    0.7931       & 31554 & 28206  &  9\farcm7 & 10\farcm5 &  5\farcm8 &  6\farcm3 & 388 & 389 & 122 & 107 & 134 & 121 \\
$18\degr < |b| < 36\degr$ E  &    0.7444       & 25816 & 30296  & 11\farcm3 & 10\farcm4 &  6\farcm8 &  6\farcm3 & 392 & 396 & 121 & 109 & 131 & 123 \\
$18\degr < |b| < 36\degr$ F  &    0.1606       & 11325 & 8536  &  7\farcm9 &  9\farcm1 &  4\farcm7 &  5\farcm4 & 134 & 107 &  30 &  19 &  34 &  24 \\
\hline
\end{tabular}
\end{table*}

\section{Satellite and bridged clusters} \label{s:sat12y}

As stated in Section~\ref{ss:prisel}, the two superselected catalogues contain a small 
number of clusters with some anomalous values of the parameters, which can be due to 
relatively rare occurrences, such as that of a very rich cluster in the neighbourhood, or to an 
unresolved close pair of sources, and so on.
A certain identification of these clusters requires more specific analyses in smaller fields with
properly suited selection values.
In the following subsections we discuss the main types of these peculiar clusters. They are reported in the final catalogue accompanied by special notes.

\subsection{Satellite clusters} \label{ss:sat}

Satellite clusters may be spurious features found in the surroundings of a rich broad cluster, 
that is, a large cluster with very many photons, typically $n \gtrsim 50$ and $R_\mathrm{max} 
\gtrsim 20\arcmin$. 
The entire 12Y-MST catalogue contains 110 sources with a maximum radius above the latter threshold, that is, 6.6\% of the entire catalogue. 
The possible satellite clusters are found to be individual minor structures that are not joined to 
the main structure because several photons may be at a distance just exceeding the separation length.

\begin{figure}[]
\vspace{0.1 cm}
\centerline{
\includegraphics[width=\columnwidth]{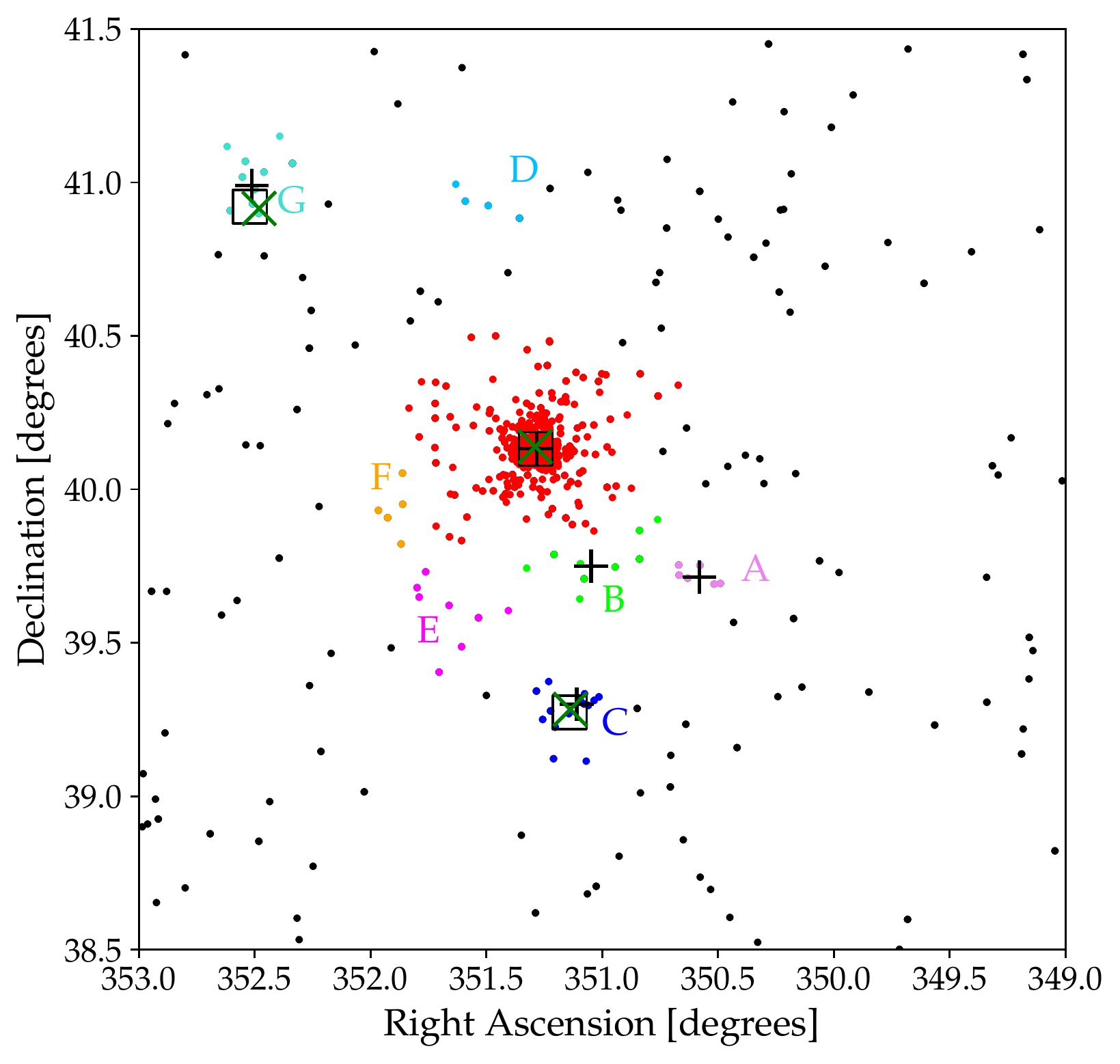}}
\caption{ $\gamma$-ray field around the very rich cluster 12Y-MST J1512-0906 (red circles), 
showing the clusters found by MST after the primary selection. 
Other coloured circles are the clusters in this region (also indicated by a letter), black 
crosses are the positions in the 12Y catalogue, large black squares are 4FGL sources, and green 
crosses correspond to the position of blazar-like objects.
Clusters D, E, and F are rejected in the secondary selection.}
\label{f:f}
\end{figure}

A good example is shown in Figure~\ref{f:f}. The figure shows the photon map in a small field around a rich 
cluster ($n = 371$), corresponding to the blazar 5BZQ J1512-0905 (PKS 1502+106, \citealt{abdo10}).
Seven other clusters, indicated by letters from A to G, are found by MST with the primary selection,
but three of them (D, E, and F) are ruled out in the secondary selection.
Cluster B, with 9 photons and a low $g$, is relatively large and lacks any concentration 
close to the centroid position, as expected for point-like sources. It can be considered a satellite 
cluster.
On the other hand, clusters C and G, with 14 and 10 photons, respectively, are clearly separated 
from the rich central cluster. Their $M$ values are higher than the threshold, as is typical of genuine sources.
Cluster A has a more uncertain status: it has only 6 photons, but a high $g = 4.3$, its distance 
from the rich cluster is larger than that of satellite B, and it is therefore likely
to be a genuine cluster, even if the possibility of a localised background fluctuation cannot 
be fully excluded.

More focused searches were performed to sort possible satellites in the catalogue by applying a
simple algorithm that selects pairs of clusters with an angular separation within the sum of their 
$R_\mathrm{max}$ values increased by 10\%.
The structures of these pairs were then analysed by applying different $\Lambda_\mathrm{cut}$ in small 
fields.
Generally, satellite clusters do not correspond to genuine $\gamma$-ray sources, but
they were not removed from the catalogue in any case. They are identified by the note `sat'.

\subsection{Halo clusters} \label{ss:hal}

Some small clusters are characterised by a median radius $R_m$ that is close to the 
average value, but a maximum radius that exceeds the corresponding typical values by far.
The simplest explanation is that one or two photons are in the surrounding background,
with a separation from the nearest photon of the cluster just below $\Lambda_\mathrm{cut}$. 
Therefore they are connected to the cluster itself.
In the regions in which the background density is higher than the average, for instance
close to the low Galactic latitude boundary, these structures are more likely detected. They are individual sources, but have an extended {halo} that is due to local background 
events.
A method for verifying the nature of the so-called {halo clusters} is to perform another
primary selection with a lower $\Lambda_\mathrm{cut}$, comparing the parameters with the
previous halo cluster and verifying whether the values of $R_\mathrm{max}$ decrease.
In the 12Y catalogue,  clusters with a possible halo structure are indicated with the
note `Hc', and their parameters are those obtained from narrow-field analysis with 
a more correct $\Lambda_\mathrm{cut}$.

\subsection{Bridged clusters} \label{ss:bri}

Bridged clusters are clusters that lie at a short angular distance and are 
connected in a unique structure by one or more background photons at angular distances 
smaller than the separation length in the field. They are rarer than satellite clusters. 
 The likelihood of finding such a structure is higher in dense fields, likely those at low Galactic 
latitude.
A simple criterion for identifying candidate bridged clusters is based on the comparison
of $R_{m}$ and $R_\mathrm{max}$: A good indicator is a value of $R_\mathrm{max}$ much 
higher than the typical value for clusters with a similar number of photons, together with a 
large $R_{m}$ because the centroid position may be located in the low-density gap between 
the two nearby features.
Another indication for the possible occurrence of a bridged structure are 
two nearby likely counterparts at an angular distance smaller than $R_\mathrm{max}$.

A rather simple method for verifying that two clusters are really connected consists of
performing another MST analysis in a small field with a typical size of 
$10\degr \times 10\degr$ or slightly larger, that is selected to avoid nearby very 
rich clusters, which can strongly reduce the mean distance between photons.
When the primary selection is iterated with a decreasing $\Lambda_\mathrm{cut}$ , 
the cluster is divided into two close structures with values of $g$ and $M$ above the 
threshold, and the bridged cluster can be considered to be resolved into its components.

An example is the cluster 12Y-MST J0843+6706, which has 22 photons, $R_m$ = 9\farcm8 and
$R_{max}$ = 15\farcm2. As we discuss in Section~\ref{s:cat12y}, only very few clusters have 
such a high $R_m$. 
A map of a small sky region of only $1\fdg5 \times 1\fdg5$ centred at the position of
this cluster is shown in Figure~\ref{f:f}. The left panel shows the cluster detected with
the original separation length: There is no central concentration around the centroid
(red cross), as expected for a point-like source, and the two 4FGL counterparts are at 
a larger distance than is generally measured for other associations.
However, several photons at $l = 148\degr$ connect two smaller structures.
The right panel of Figure~\ref{f:f} shows the clusters obtained by computing the MST in a 
field that extends 20\degr\ in longitude and 10\degr\ in latitude, with 
$\Lambda_\mathrm{cut} = 0.5$.
The two bridging photons are excluded, and there are now two 
clusters of 10 photons each, whose centroids are correctly located and nearly coincident 
with the 4FGL positions.

In the 12Y catalogue, clusters with a possible bridge structure are indicated with the
note `Bc', followed by a letter to relate the uncoupled clusters. 
Their reported properties refer to the resolved components.

In some cases, however, only one of the resolved clusters has parameters that are high enough to exceed
the secondary selection thresholds. Only the good cluster is included in the catalogue.
In a few rare cases, a bridge or a halo pattern in a cluster cannot be easily
recognised unless the components are separated by applying a very short $\Lambda_\mathrm{cut}$.
These cases are indicated with the note `HBc'.

The rich cluster 12Y-MST J0348-2750 with $n = 353$, 
$R_m$ = 5\farcm22 and $R_\mathrm{max}$ = 51\farcm54 is an interesting peculiar case.
This maximum radius is the largest in the catalogue. Clusters that are even richer, up to 1000 
photons and more, have lower values of $R_\mathrm{max}$. The median radius is instead comparable to those of 
similar clusters.
This anomaly suggests that it may be an unresolved system.
A MST search in a small field with $\Lambda_\mathrm{cut}$ reduced to 0.4 resolved three clusters.
The richest has 304 photons and $R_\mathrm{max} = 22\farcm86$, which agrees very well with 
other typical values. The other two have only 7 and 5 photons, but high $g$ and magnitudes 
$M$ that exceed the secondary selection threshold.
In particular, the cluster with 7 photons corresponds to a 4FGL source and is nearly 
coincident with a blazar counterpart. The third closer and poorer cluster might indeed be a 
satellite.

\begin{figure}[]
\vspace{0.1 cm}
\centerline{
\includegraphics[width=\columnwidth]{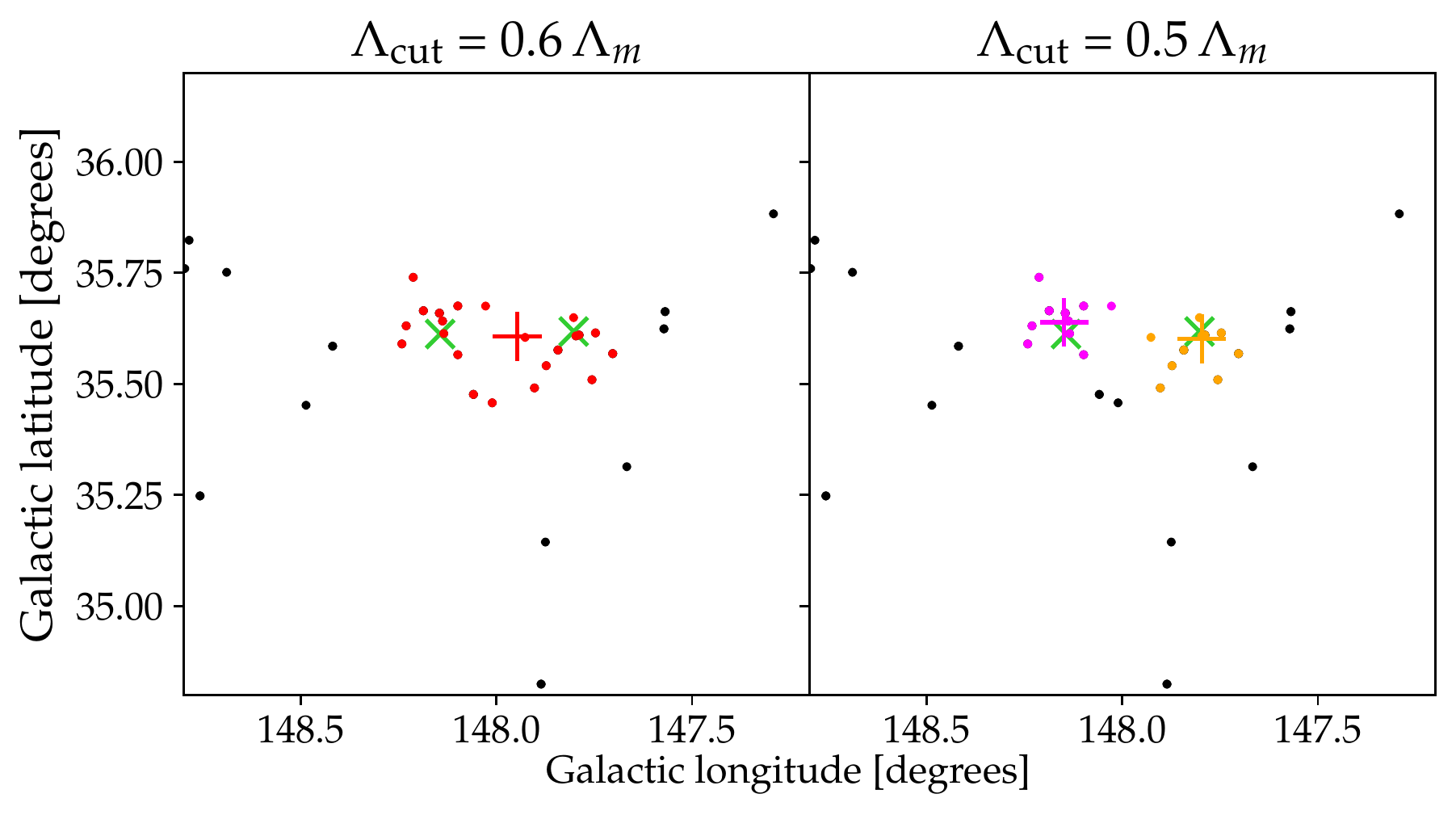}}
\caption{ Photon map of the sky region around cluster 12Y-MST J0843+6706. Left 
panel: 
Black points are background photons, and red points are photons in the cluster after the 
selection with $\Lambda_\mathrm{cut} = 0.6\,\Lambda_{m}$. The red cross marks the centroid 
position, and the two green crosses are the positions of 4FGL sources.
Right panel: Same field after the analysis with $\Lambda_\mathrm{cut} = 0.5\,\Lambda_{m}$. The magenta and orange points are the photons of the two resulting clusters.}
\label{f:g}
\end{figure}

\begin{figure}[]
\vspace{0.1 cm}
\centerline{
\includegraphics[width=\columnwidth]{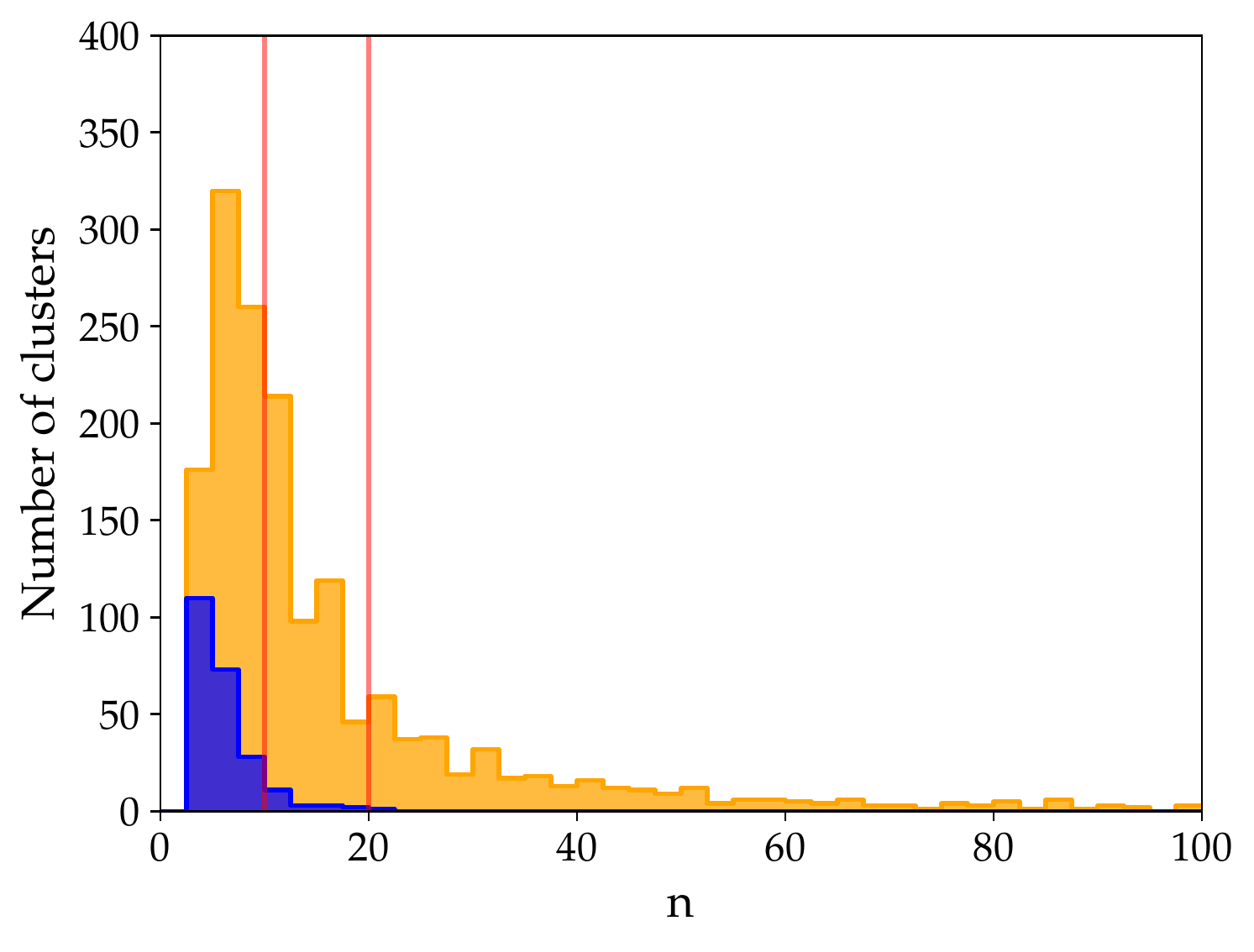}}
\caption{Orange histogram: Photon number in the clusters of the 12Y-MST 
catalogue. Violet histogram: Clusters without a corresponding source
in the 4FGL catalogue.
The horizontal range is limited to 100 to clearly show the distribution for the clusters 
with low photon numbers.
The vertical lines mark the median and the 75th
percentile.
}
\label{f:b}
\end{figure}

\begin{figure}[]
\vspace{0.1 cm}
\centerline{
\includegraphics[width=\columnwidth]{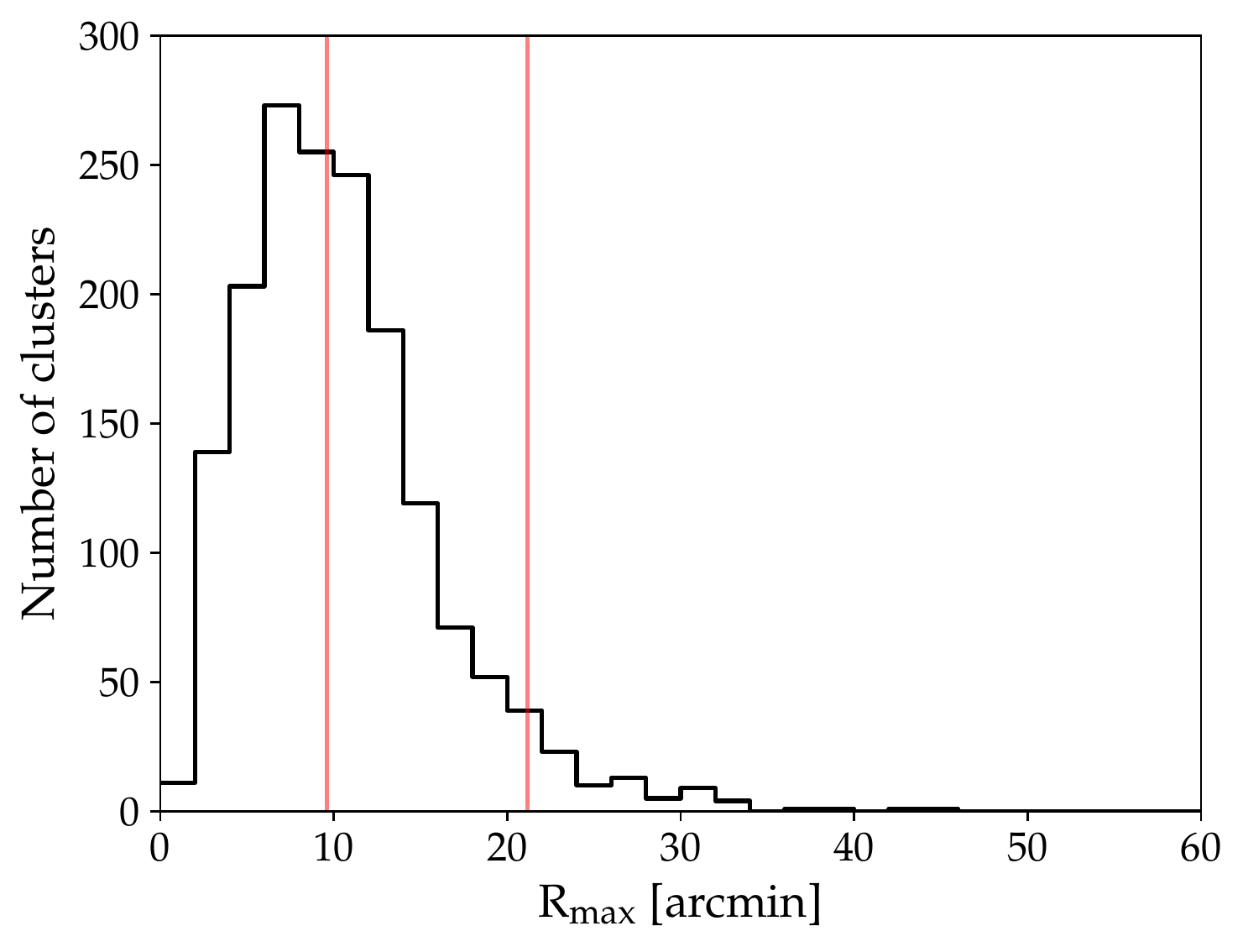}}
\caption{Histogram of the maximum radius of the clusters of the 12Y-MSTs catalogue.
The vertical lines mark the median and the 95th 
percentile.
}
\label{f:c}
\end{figure}

\begin{figure}[]
\vspace{0.1 cm}
\centerline{
\includegraphics[width=\columnwidth]{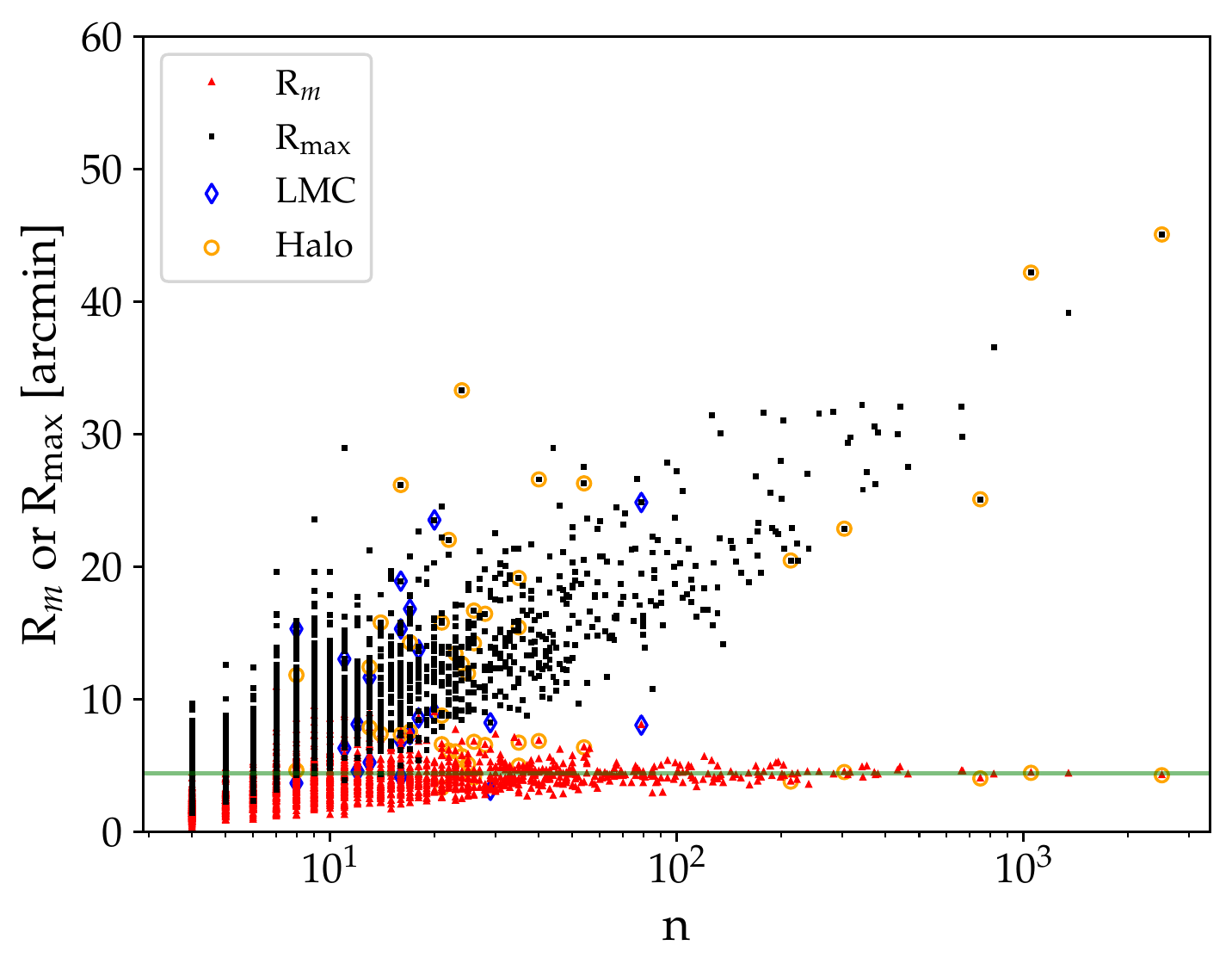}}
\caption{Bivariate distributions of the photon number vs the radius delimiting
the circle containing 50\% of the photons, $R_m$ (red triangles), and maximum radius 
(black squares) of the clusters in the 12Y-MSTs catalogue.
Clusters with large haloes and in those in the LMC region are indicated.
The horizontal line corresponds to the constant fit of the median radius of
clusters with a number higher than 90. 
}
\label{f:d}
\end{figure}

\section{12Y-MST catalogue} \label{s:cat12y}

\subsection{General cluster properties} \label{ss:proper}

In the following we describe some general properties of the selected clusters in the
main 12Y catalogue and the distribution of their parameters, while those of the 12Yw
addendum are presented in Section~\ref{s:12yw}.
The first property is the number of photons (Figure~\ref{f:b}). Because of the primary 
selection, this number is equal to or greater than 4.
The highest number of photons found in a single cluster is 2503, and this corresponds
to the well-known BL Lac object Mk~421.
However, the median of the distribution is at $n_\mathrm{md} = 10,$ and only 25\% of 
clusters have $n > 20$.

We showed that a useful parameter to describe the spatial structure of clusters is   
$R_\mathrm{max}$.
This radius is generally larger than the instrumental PSF, and in a few very rich clusters 
can exceed 30\arcmin.
The histogram of $R_\mathrm{max}$ is reported in Figure~\ref{f:c}. As expected, the large 
majority of clusters has a maximum radius smaller than 20\arcmin, and the median value of 
the distribution is 9\farcm6.
As mentioned above, it is useful to perform a further analysis for the largest clusters to 
verify whether they are bridged or halo clusters, although it cannot be excluded that some of 
them correspond to extended structures.
Extended features are generally found close to the Galactic equator, but a few of them 
are in the LMC region.

The behaviour of $R_m$ is different because it results from the distributions
shown in Figure~\ref{f:d}.
For rich clusters, $R_m$ approaches a value equal to 4\farcm4, estimated by the 
mean for clusters with $n > 90$, with a remarkably low dispersion. The maximum radius 
increases slowly with $n$ up to values higher than $R_m$  by about one order of magnitude.
Some outlier points correspond to halo clusters, as discussed
above.

\subsection{12Y-MSTw sample}
\label{s:12yw}

As discussed in Section~\ref{ss:secsel},  a different superselection that applied weaker
thresholds than were used for the standard `strong' 12Y-MST resulted in
 the additional 12Y-MSTw sample of 224 candidate sources.
All these clusters have fewer than ten photons, and only ten clusters have $g$ values 
higher than 4 and none is higher than 5.
Therefore the possibility of including spurious clusters in this sample is expected to be 
higher than in the standard catalogue.
Furthermore, four possible weak satellites of three rich clusters in the 12Y were found, 
indicated as usual by the note `sat'.
The search for correspondences with other catalogues resulted in associations for about half
of the sample. This confirms the usefulness of the weak selection.

\section{Correspondences between 12Y-MST and other catalogues}
\label{s:corr12y}

\subsection{3FHL}
\label{ss:corr3FH}

The third \FLL Catalogue of High-Energy Sources, 3FHL \citep{ajello17},
contains 1556 entries corresponding to sources detected at energies higher than 10 GeV, 986 of which are at Galactic latitudes higher than 20\degr. 
In this subset, four sources are classified as extended: three of them are in the LMC
region, and the last is centred on the active galaxy Fornax A (NGC 1316) and its
surroundings.
The observation time window of this catalogue is limited to the first seven years of observation,
and more recent versions are not yet available.
A search for correspondences with 12Y-MST clusters within an angular separation of 9\arcmin\
gives 945 matches (95.8\% of the 3FHL sample);
only one source is at an angular separation higher than 6\arcmin.
Considering that the 12Y catalogue has several clusters in the LMC region and another two
in the Fornax A region (but their centroid coordinates do not match those of
the 3FHL catalogue within the adopted distance), the number of effective correspondences
might be increased to 948.

Eleven more associations are found by searching for  correspondences between 3FHL and the 
12-MSTw sample.
A further analysis of clusters found in the primary selection whose parameters
were below the secondary selection thresholds gave 16 more matches.
As a conclusion, only 14 3FHL sources are {not} detected by MST algorithm, which is about 1.5\% of the full catalogue.

This number of correspondences is only slightly higher than the one found for the 9Y 
\citep{campana18}, 
as expected from the longer exposure of the current version.
However, ten of the unassociated 3FHL sources are at 
Galactic latitudes lower than 30\degr, where the cluster search is less sensitive
given the higher local background.

\subsection{4FGL-DR2 and other catalogues}
\label{ss:corr4F}

The 4FGL-DR2 catalogue is the most recent version of the all sky $\gamma$-ray sources 
in the energy range from 50 MeV to 1 TeV released by the \FLL collaboration 
\citep{Abdollahi20,Ballet20}.
The DR2 version is the most recent version on the \FLL website. It is based on data 
acquired in ten years of activity and has 5788 entries, 3237 of which are at $|b| > 20\degr$.
A search for positional correspondences between our catalogues and 4FGL within an 
angular separation $\Delta \leq 9\arcmin$ resulted in 1430
associations (86\%) for the 12Y-MST sample, with a mean angular distance between the 
centroid coordinates of the clusters and the 4FGL source position of $\Delta_m =$ 1\farcm5.
The histogram of angular distances of these associations is reported in Figure~\ref{f:e}.
Only 
13 associations are at values of $\Delta$ higher than 6\arcmin\ (only 6 of them above
7\arcmin).
Their nature is discussed in Section~\ref{sss:corr4Fld}.

The number of chance random associations can be evaluated by considering the ratio
of the solid angles covered by all the clusters in 12Y, assuming for each of them
an angular radius equal to the radius used in the matching, and the solid angle of the whole
explored sky.
This ratio is equal to 4.328 $\times 10^{-3}$ , and thus the expected number of random 
associations is 14. This is lower than what we found by two orders of magnitude
and therefore confirms that almost all of the associations can be considered genuine.

The 12Y-MSTw and the 4FGL-DR2 contain 96 associations (about 43\% of this sample)
within the same matching distance. Only 6 sources lie above 6\arcmin. 
This notable number of further correspondences confirms the validity of the cluster detection.
The $\Delta_m$ value is equal to 2\farcm6, 
which is far lower than the instrumental PSF. This implies a good accuracy of the MST positional 
estimates.

The remaining set of 234 12Y-MST clusters that are not associated with 4FGL-DR2 sources was analysed. We found 3 other counterparts with the 3FGL catalogue \citep{acero15}.

Several candidate sources in this sample are in the LMC region where the local
photon density is much higher than in the typical high-latitude sky. They might correspond to extended structures.
One of them corresponds to a 3FGL source, while another two clusters confirm with
a high significance the detection of the two supernovae remnants (SNRs) 
N 49B and N 63A 
\citep{campana18b} found in the analysis of the nine-year LAT sky.
A full analysis of the complex LMC region is beyond the scope of this paper
and will be presented in a forthcoming paper.

Excluding these sources, we find that {all} the 12Y candidate sources with 
$n > 18$ are associated with 4FGL-DR2 sources. Only 26 of the 588 clusters 
with $9 \leq n \leq 18$ do not have such correspondence.

Pulsars are another relevant class of high-energy sources, but only a small fraction 
of them is at high Galactic latitude.
Therefore only few pulsars are expected to be in our catalogue.
Only the millisecond pulsar PSR J0843+67 was reported in the 9Y-MST, and it is 
confirmed in 12Y.
Moreover, PSR J2234+0944 is associated with a cluster in the 12Y-MSTw sample.
Both sources are in the 4FGL-DR2 and in previous reports of the \FLL collaboration.

\subsubsection{Associations at $\Delta > 6\arcmin$}
\label{sss:corr4Fld}

\begin{figure}[ht]
\vspace{0.1 cm}
\centerline{
\includegraphics[width=\columnwidth]{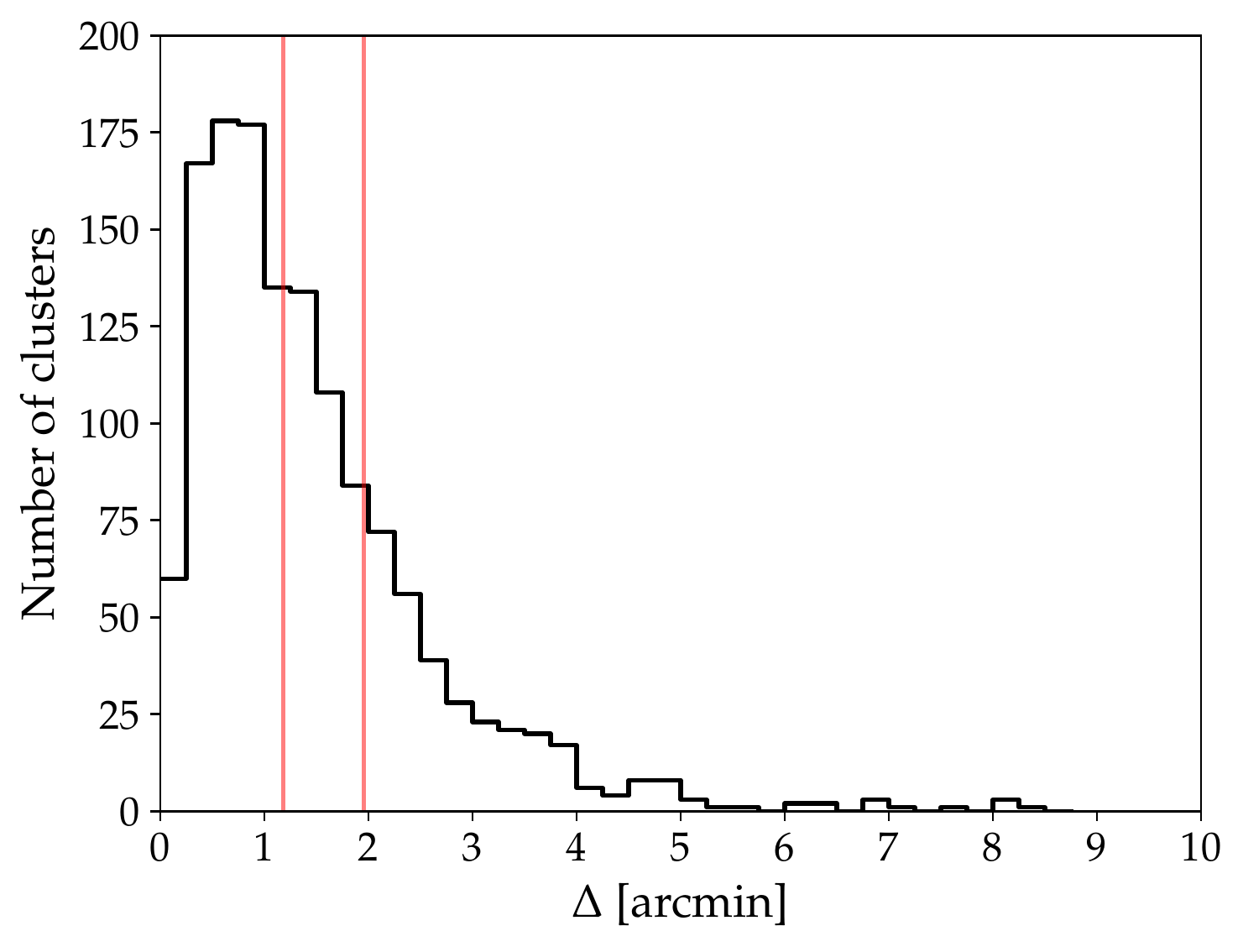}}
\caption{Histogram of the angular separation between 12Y clusters and 4FGL counterparts.
The red vertical lines mark the median and the 75th percentile.}
\label{f:e}
\end{figure}

The histogram in Figure~\ref{f:e} shows a quite small subset of 12Y-MST 
clusters that match 4FGL sources with an angular separation in the range 
$6\arcmin \leq \Delta \leq 9\arcmin$.
This subset contains 13 clusters, that is, only 1\% of the associated clusters, and
therefore it cannot be excluded that they are the result of possible random proximity.
The number of random associations with an angular distance in the
previous range can be evaluated by means of the corresponding solid angle. We obtain 
an estimate of 8, which is significantly smaller than the associations we found.

Moreover, the source coordinates in the 4FGL catalogue are
obtained from the analysis in a much broader energy band that extends to about 50 MeV. It is therefore possible that positional discrepancies higher than
expected  are found for a few sources.

The nature of the sources in this subset was accurately investigated, in particular for
the search of possible counterparts.
We found that ten of these clusters are associated with known blazars or candidates.
Another cluster is at a distance of 1\farcm8 from the position of a 3FGL source, while the others, 
all at large distances, are without counterpart.

\subsection{TeV catalogues}
\label{ss:tev}

The comparison of 12Y with available catalogues of sources detected at TeV energies 
requires particular attention because the sky coverage in this very high energy band 
is scarcely uniform.
A catalogue of sources detected by many TeV observatories is available at the website of the University
of Chicago\footnote{version 3.400 at \url{http://tevcat.uchicago.edu/}}.
A nearly complete list of known TeV sources, referred to as TeGeVcat \citep{carosi15}, is 
available at the ASI-SSDC website\footnote{\url{https://www.ssdc.asi.it/tgevcat/}}.
This catalogue is updated up to December 2018, and therefore some other 
recently reported \citep{acciari20} were added.
Another list (compiled by M. Mori) is available at the Ritsumeikan 
University\footnote{\url{http://www.ritsumei.ac.jp/~morim/TeV-catalog/index.html}}.

By combining the information provided by these catalogues, a list containing 79 
entries at $|b| > 20\degr$ was obtained, including some GRBs.
When GRBs not detected by LAT above 10 GeV or outside the field of view are excluded, the useful
sample for the comparison with 12Y is reduced to 75. Four sources lie in the LMC region.

We found 65 correspondences in the 12Y catalogue. Thirty-eight of these clusters have more than 
50 photons and 51 clusters have more than 30 photons, but there is also a cluster with only 6 photons.
The 12Y catalogue therefore includes a very large fraction of the TeV sources detected in 
the sky with $|b| > 20\degr$, and therefore it may be profitably used to select several 
possible targets for future pointed TeV observations.

\subsection{9Y-MST}
\label{ss:corr9y}

The comparison of the new 12Y-MST and the previous 9Y-MST catalogue is useful to 
evaluate the efficiency of the MST algorithm in detecting low-significance features and 
in determining their stability.
We therefore searched for the correspondence of cluster centroid coordinates in the two 
catalogues within an angular distance of 10\arcmin\  and found that 1206 of the 1342 9Y 
catalogue sources correspond to 12Y clusters (the mean distance is 0\farcm8, the maximum 
distance is 8\arcmin, but only 3 clusters have a distance higher than 6\arcmin). 
The remaining 137 sources are without correspondence.
Many of these unassociated 9Y clusters were rejected by the severe thresholds of 
the secondary selection on the 12Y dataset, however. Twenty-four of them are in fact included in the catalogue of the weak selection. Another 49 clusters were found in the primary selection, but their parameters were too low 
to exceed any threshold, and 3 clusters are in the LMC region, where a safe correspondence
is biased by the high photon density.
This means that the number of 9Y clusters that are `lost' is 61, corresponding to 4.6\% 
of the total sources in this catalogue.
This value can be compared with the percentage of 3FGL sources that are not included
in the 4FGL within a radius of 15\arcmin, which is about 13\%.
Forty of the 9Y-lost clusters have 4 or 5 photons, and the 6 clusters with $n > 9$ have $g$ values 
lower than 2.6, and in one case, as low as 1.8.
Therefore the parameters of these clusters were generally just above the selection thresholds.

\section{New high-energy detections of blazars and blazar candidates}\label{s:newblz}

It is well established that the large majority of extragalactic $\gamma$-ray sources are active 
galactic nuclei of blazar type \citep{massaro15}.
We therefore searched for a possible blazar association in the subsample of newly detected candidate sources
in the 12Y and 12Yw catalogues using the 5BZCAT catalogue \citep[][the most complete available 
blazar sample, although not updated with recent findings]{massaro14,massaro15b}.
This search resulted in 833 associations with 12Y clusters within a distance of 6\arcmin, and other 8 
associations within 8\arcmin. 
There are therefore 107 (or 113 within the latter distance) new blazar counterparts, 17
of which do not have a corresponding source in the 4FGL or in the 9Y catalogues. These are therefore new 
detections.
The same search for the 128 unassociated clusters in the 12Yw resulted in 7 more new associations with 
5BZCAT objects.
In total, 24 new detections of blazars above 10 GeV are reported, 
and their 5BZCAT identifications are given in the notes.

In the past years, several papers 
have reported lists of blazar candidates that were identified by applying different selection 
criteria based on the occurrence of some of their typical properties.
We investigated whether some of these objects might be associated with 12Y clusters without correspondences 
in the 4FGL catalogue because possible known counterparts are already reported there. 
This search in the sample of 234 clusters provided 34 possible associations with candidate blazars.
The breakdown is as follows: 
5 in KDEBLLACS\footnote{Kernel Density Estimation BL Lacs, \cite{dabrusco19}}, 2 with WIBRaLS2\footnote{\emph{WISE} Blazar-like Radio-Loud Sources, \cite{dabrusco19}}, one from ROXA\footnote{Radio-Optical-X-ray catalog, \cite{turriziani07}}, 16 in 2WHSP\footnote{\emph{WISE} High Synchroton Peaked blazars, \cite{chang17}}, and 10 (one cluster 
with a double association in the Fornax region) in CRATES\footnote{Combined Radio All-Sky Targeted Eight GHz Survey, \cite{healey07}}.
A similar analysis of the additional 12Yw list resulted in 8 more associations (2 in KDEBLLACS, 3 in 2WHSP,
and 3 in CRATES). The total number of detections of confirmed blazars and candidates 
thus increases to 42.
Again, all these possible counterparts are reported in the catalogues as notes.

As an example, Table~\ref{t:newsrc} reports the first 30 12Y catalogue entries without association to 4FGL sources, ordered by right ascension. The full catalogue is provided in electronic format through the CDS archive.

\section{Summary and discussion}
\label{s:sumcon}

The 9Y-MST catalogue of $\gamma$-ray candidate sources, based on the first nine years of Fermi-LAT data, 
has been extended to include three more years of observations, thus producing the new 12Y-MST catalogue.
Pass 8 events were selected at energies higher than 10 GeV and Galactic latitude $|b| > 20\degr$,
and the MST algorithm was applied to detect photon clusters. 
As for the 9Y, after the primary selection, severe threshold values were adopted to reduce 
the possibility of spurious detections due to local background fluctuations.
A new catalogue of 1664 candidate sources was obtained. These are 322 (24\%) more than in the 9Y catalogue.

By applying weaker selection criteria, an additional sample 
of clusters was selected, with a lower significance, but useful to achieve a richer sample 
of possible high-energy sources.
This 12Y-MSTw catalogue consists of 224 additional clusters to the 12Y for a total  of 1888 
candidate sources. This total number is about twice those reported in the 3FHL catalogue in 
same sky regions and energy band.

About 80\% of the 12Y clusters have a very good positional correspondence with sources reported in the
recent 4FGL-DR2 catalogue of the \FLL collaboration. The remaining 20\%, or at least a
significant fraction of them, are expected to be the high-energy counterpart of blazar-like AGNs.
Moreover, almost all the confirmed TeV sources in the investigated regions of the sky are included in the 12Y. 
As a consequence, the catalogue can be used to derive a sample of possible targets for current 
and future very high energy observatories.

A search for new possible blazar candidates was carried out within a region centred at the 12Y 
or 12Yw cluster centroid coordinates. The region had a radius of 6\arcmin\  . The search was based on possible optical 
or IR counterparts of radio sources, when present, or of quasars or candidates reported 
in large databases.
As a result, 24 new detections of blazars that were reported in the 5BZCAT were obtained and 
another 42  possible associations with blazar candidates.
Moreover, possible interesting radio and IR sources are located within a few arcminutes
from the positions of several cluster centroids. Further observational studies are required to
confirm them as reliable counterparts.

\begin{acknowledgements}

We acknowledge use of archival Fermi data. We made large use of the online version of 
the Roma-BZCAT and of the scientific tools developed at the ASI Science Data Center (ASDC),
of the final release of 6dFGS archive,
of the Sloan Digital Sky Survey (SDSS) archive, of the NED database and other astronomical 
catalogues distributed in digital form (Vizier and Simbad) at Centre de Dates astronomiques de 
Strasbourg (CDS) at the Louis Pasteur University.
This research has made use the TeVCat online source catalog (\url{http://tevcat.uchicago.edu}).
\end{acknowledgements}

\bibliographystyle{aa}
\bibliography{bibliography} 

\onecolumn

\begin{landscape}
\begin{longtable}{lrrrrrrrrrrrrr}
\caption{\label{t:newsrc}  The first 30 entries in the 12Y-MST catalogue not associated with any 4FGL \emph{Fermi}-LAT source. 
Column $N$ lists the number of photons in the MST cluster, $g$ its clustering degree, and $M = Ng$ its magnitude. See main text for details and for the meaning  notes.
}\\ \hline\hline
Name                & RA (J2000)       & Dec (J2000)     & $l$        & $b$        &  $N$   & $g$     & $M$        &  $R_m$    & $R_\mathrm{max}$   & Notes \\
                        & \degr       & \degr     & \degr        & \degr        &     &      &         &    \arcmin  &  \arcmin  &  \\
\hline
\endfirsthead
\caption{continued.}\\\hline\hline
Name                & RA (J2000)       & Dec (J2000)     & $l$        & $b$        &  $N$   & $g$     & $M$        &  $R_m$    & $R_\mathrm{max}$   & Notes \\
                        & \degr       & \degr     & \degr        & \degr        &     &      &         &  \arcmin    & \arcmin   &  \\
\hline
\endhead
\hline
\endfoot
  12Y-MST J0000$-$0216 & 0.104 & $-$2.272 & 94.596 & $-$62.303 & 4 & 4.292 & 17.168 & 2.82 & 5.94 & \\
  12Y-MST J0003$-$0705 & 0.825 & $-$7.093 & 91.053 & $-$66.936 & 4 & 3.601 & 14.405 & 2.10 & 7.50 & \\
  12Y-MST J0009$+$3655 & 2.411 & 36.918 & 113.710 & $-$25.204 & 4 & 4.291 & 17.163 & 0.72 & 6.06 & \\
  12Y-MST J0009$+$2810 & 2.438 & 28.167 & 111.868 & $-$33.808 & 6 & 3.408 & 20.446 & 4.74 & 7.74 & KDEBLLACS\\
  12Y-MST J0010$+$2422 & 2.599 & 24.378 & 111.124 & $-$37.550 & 4 & 5.906 & 23.622 & 1.38 & 2.70 & \\
  12Y-MST J0014$-$3223 & 3.573 & $-$32.393 & 357.540 & $-$80.378 & 4 & 4.676 & 18.706 & 4.44 & 6.84 & \\
  12Y-MST J0019$-$0922 & 4.840 & $-$9.369 & 98.329 & $-$70.692 & 6 & 3.066 & 18.397 & 5.04 & 12.42 & \\
  12Y-MST J0021$-$2340 & 5.463 & $-$23.682 & 58.593 & $-$82.485 & 4 & 5.852 & 23.409 & 1.68 & 6.48 & \\
  12Y-MST J0026$-$2003 & 6.612 & $-$20.051 & 82.661 & $-$80.900 & 7 & 2.733 & 19.134 & 6.36 & 13.38 & \\
  12Y-MST J0030$-$4959 & 7.537 & $-$49.992 & 311.631 & $-$66.776 & 4 & 3.799 & 15.196 & 3.36 & 4.80 & \\
  12Y-MST J0032$-$1908 & 8.167 & $-$19.145 & 93.580 & $-$80.927 & 6 & 3.200 & 19.201 & 5.22 & 8.16 & sat[12Y-MST J0032$-$1908]\\
  12Y-MST J0032$-$4723 & 8.174 & $-$47.390 & 311.979 & $-$69.407 & 7 & 2.859 & 20.011 & 7.50 & 16.44 & 2WHSP J003222.5$-$47253\\
  12Y-MST J0035$-$2638 & 8.846 & $-$26.646 & 39.700 & $-$86.388 & 5 & 4.340 & 21.698 & 2.94 & 8.28 & \\
  12Y-MST J0044$+$1212 & 11.194 & 12.210 & 120.365 & $-$50.628 & 6 & 3.073 & 18.440 & 3.24 & 7.20 & sat[12Y-MST J0045$+$1217]\\
  12Y-MST J0046$+$2147 & 11.608 & 21.798 & 121.391 & $-$41.058 & 4 & 4.048 & 16.192 & 2.28 & 3.30 & \\
  12Y-MST J0058$-$1433 & 14.562 & $-$14.55 & 130.459 & $-$77.322 & 4 & 4.533 & 18.131 & 0.90 & 8.04 & 2WHSP J005813.7$-$14292\\
  12Y-MST J0111$-$4902 & 17.877 & $-$49.034 & 294.220 & $-$67.753 & 4 & 4.770 & 19.082 & 1.92 & 5.46 & \\
  12Y-MST J0119$-$1714 & 19.775 & $-$17.234 & 157.221 & $-$78.220 & 4 & 4.143 & 16.571 & 3.36 & 7.86 & \\
  12Y-MST J0119$-$2738 & 19.961 & $-$27.640 & 219.196 & $-$83.674 & 7 & 2.701 & 18.910 & 10.92 & 19.62 & \\
  12Y-MST J0121$-$7216 & 20.309 & $-$72.270 & 299.749 & $-$44.674 & 6 & 5.091 & 30.549 & 2.82 & 3.60 & \\
  12Y-MST J0121$+$3807 & 20.451 & 38.123 & 129.483 & $-$24.362 & 4 & 6.579 & 26.318 & 1.68 & 2.52 & \\
  12Y-MST J0125$-$2723 & 21.279 & $-$27.395 & 216.898 & $-$82.512 & 4 & 3.913 & 15.653 & 4.02 & 5.94 & \\
  12Y-MST J0127$+$1737 & 21.984 & 17.627 & 135.140 & $-$44.378 & 4 & 6.204 & 24.815 & 1.38 & 3.30 & KDEBLLACS\\
  12Y-MST J0135$+$3834 & 23.911 & 38.573 & 132.337 & $-$23.490 & 7 & 3.243 & 22.703 & 4.32 & 13.32 & \\
  12Y-MST J0140$+$0721 & 25.182 & 7.362 & 143.778 & $-$53.503 & 4 & 5.124 & 20.495 & 2.10 & 3.72 & \\
  12Y-MST J0145$+$1000 & 26.386 & 10.007 & 144.218 & $-$50.616 & 4 & 4.079 & 16.314 & 1.80 & 5.34 & \\
  12Y-MST J0146$-$0244 & 26.512 & $-$2.745 & 153.449 & $-$62.336 & 4 & 3.822 & 15.286 & 1.62 & 7.26 & 2WHSP J014620.2$-$02425\\
  12Y-MST J0153$-$1107 & 28.264 & $-$11.127 & 168.047 & $-$68.417 & 5 & 4.055 & 20.277 & 3.06 & 7.80 & 2WHSP J015313.1$-$11062\\
  12Y-MST J0154$-$6605 & 28.681 & $-$66.092 & 293.074 & $-$49.807 & 4 & 4.599 & 18.397 & 1.56 & 4.14 & \\
  12Y-MST J0157$-$3236 & 29.252 & $-$32.604 & 237.965 & $-$74.788 & 4 & 4.185 & 16.741 & 3.24 & 4.62 & 2WHSP J015700.6$-$32352\\
\end{longtable}
\end{landscape}

\end{document}